\documentclass[twocolumn,prl, amsmath,amssymb]{revtex4}

\usepackage{graphicx}
\usepackage{bm, color}

\newcommand{\be}{\begin{equation}}
\newcommand{\ee}{\end{equation}}

\newcommand{\vrr}{{\bf{r}}}

\newcommand{\vB}{{\bf{B}}}
\newcommand{\vH}{{\bf{H}}}

\newcommand{\vS}{{\bf{S}}}

\newcommand{\cH}{{\cal H}}

\newcommand{\si}{\sigma}

\newcommand{\eps}{\epsilon}

\newcommand{\uparr}{\uparrow}
\newcommand{\dwnarr}{\downarrow}

\newcommand{\la}{\langle}
\newcommand{\ra}{\rangle}

\newcommand{\idos}{\left. \left( \! \frac{dn}{d\eps} \! \right) \!\! \right.^{-1}}

\begin{document}

\title{Spin polarization of half-quantum vortex in systems with equal spin pairing.}

\author{Victor Vakaryuk}
\email[]{vakaryuk@uiuc.edu}
\author{Anthony J.~Leggett}

\affiliation{Department of Physics, University of Illinois, Urbana, Illinois 61801, USA}

\date{\today}

\begin{abstract}
We present a variational analysis for a half-quantum vortex (HQV) in the equal-spin-pairing superfluid state which, under suitable conditions, is believed to be realized in $\rm Sr_2RuO_4$ and $^3$He-A. 
Our approach is based on a description of the HQV in terms of a BCS-like wave function with a spin-dependent boost. We predict a novel feature: the HQV, if stable, should be accompanied by a non-zero spin polarization. Such a spin polarization would exist in addition to the one induced by the Zeeman coupling to  the external  field  and hence may serve as an indicator in experimental search for HQV.
\end{abstract}


\maketitle

When $^3$He is liquified and cooled into the millidegree regime it enters a new phase which has been proved to possess a spin triplet paired condensate. There also exists by now a growing body of experimental evidence that  $\rm Sr_2RuO_4$ below 1.5 K is a spin triplet superconductor \cite{Mackenzie:2003}. This implies the possibility 
 of many interesting phenomena not expected in systems with spin singlet pairing. One of them is the existence of half-quantum vortices (HQV's) in the equal-spin-pairing  (ESP) state of the spin triplet condensate \cite{Volovik:1976, Cross:1977}.

The pairing symmetry for $^3$He is well established and the so-called A phase is confidently believed to realize an ESP spin triplet state    \cite{Leggett:1975p283, Vollhardt:1990}. While there is, to the best of our knowledge, no unambiguous observation of HQV in $^3$He-A \cite{yamashita:2008}, there is a strong theoretical argument in favor of their existence at least under some assumptions on the geometry of the experiment. On the contrary, the pairing state of $\rm Sr_2RuO_4$ is currently  poorly understood and the observation (or not) of HQV, along with other experimental information, would facilitate identification of the underlying pairing symmetry. In addition, there is a significant interest in HQV for topological quantum computing \footnote{see e.g.~D.~A.~Ivanov, Phys.~Rev.~Lett.~{\bf 86}, 268 (2001).}. 

It should be noted right away that the identification of the superconducting phase of $\rm Sr_2RuO_4$
with an ESP state does not by itself guarantee thermodynamic stability of HQV in this compound. Even under the assumption of negligible spin-orbit coupling, the kinetic energy of unscreened spin currents which accompany HQV disfavors its formation vis-\`a-vis the formation of a regular vortex where electromagnetic currents are screened over the length of the penetration depth. Such an unfavorable energy balance can be avoided by limiting the sample size to a few microns \cite{Chung:2007}. This however further complicates the experimental detection of HQV in $\rm Sr_2RuO_4$.
 
One of the most direct ways for detection of HQV is to look for  spin currents which circulate around it.  The usual techniques for spin current detection are based on the accumulation of spin and their straightforward application to this situation seems to be difficult. One can however use a fact that spin currents generate electric field. A very rough conservative estimate shows that for a ring of size 1$\mu$ the quadrupole electric field generated by the spin currents of HQV will create a potential difference of 1nV across the ring -- quite small, but not beyond the capabilities of current experimental techniques. 

There is also a possibility  of detecting HQV by looking for specific features in the magnetization curves of small  rings made of $\rm Sr_2RuO_4$. Experiments in this direction are currently underway in the  Budakian group at UIUC.

In this paper we suggest an apparently new effect which may be utilized for the detection of HQV both in $\rm Sr_2RuO_4$ and in $^3$He-A. The effect consists in the presence of an effective Zeeman field in the HQV state of the ESP condensate. In thermodynamic equilibrium such an effective Zeeman field will produce a non-zero spin polarization in addition to that created by external fields. In particular, such a spin polarization would exist even in the absence of external Zeeman coupling provided the condensate is in the HQV state. At the same time this field would not exist in a normal vortex state thus allowing one to distinguish between the two. For a 1$\mu$ ring of $\rm Sr_2RuO_4$ the magnitude of the effective Zeeman field is about 10 Gauss and, taking the spin susceptibility to be of order $10^{-3} \rm emu/mol$ \cite{Mackenzie:2003},   the spin polarization produced by such a field can be seen in $T_1$ or even Knight shift measurements.

 We start by noticing that in the ESP state the Cooper pair is always in a linear superposition of states in which both spins in the pair are either aligned (``up'') or antialigned (``down'') with a given direction in space. The corresponding many-body wave function for a system with $N/2$ pairs which are condensed into the same two-particle state characterized by functions  $\varphi_\uparr$ and $\varphi_\dwnarr$ can be written as
 \be
 \begin{split}
 	\Psi_{\rm ESP}
	=
	{\cal A}
	\Big\{
	\big[
	\varphi_{\uparr}(\vrr_1, \vrr_2)  \, | \!\! \uparr \uparr \ra
	+
	\varphi_{ \dwnarr}(\vrr_1, \vrr_2) \, | \! \! \dwnarr \dwnarr \ra
	\big]
	\ldots
	\\
	\big[
	\varphi_{ \uparr}(\vrr_{N-1}, \vrr_N) \, | \! \! \uparr \uparr \ra
	+
	\varphi_{\dwnarr}(\vrr_{N-1}, \vrr_N) \, | \! \! \dwnarr \dwnarr \ra
	\big]
	\Big\},
\end{split}
\ee
 where $\cal A$ is the antisymmetrization operator with respect to particles' coordinates $\vrr_i$ and spins.
 In the weak coupling limit, provided the pairing interaction conserves spin, the up and the down spin particles can be considered as independent subsystems.  In this case the HQV state of the ESP condensate has a simple physical interpretation: It is a state in which the two spin systems have different winding numbers, i.e.~accommodate different number of vortices. 

To avoid complications related to the presence of the vortex core we specialize to an annular geometry.  Let $R$ be the radius of the annulus and $d$ be the wall thickness; it will be assumed that $d/R \ll 1$ so that effects of order $d/R$ or higher can be ignored. Then specializing to the zero-temperature case and choosing  the spin axis  along the symmetry axis of the annulus,  a conceptually simple ansatz for the HQV state of the condensate is 
\be
	\Psi_{\text{HQV}}
	= 
	\exp 
	{\Big\{ 
	 \frac{i \ell_\uparr}{2} 
	\sum_{i =\uparrow} \theta_i 
	+
	  \frac{i \ell_\dwnarr}{2}
	 \sum_{i =\downarrow}  \theta_i 
	\Big	\}}
	 \,\, 
	\Psi_{\rm ESP}, 
\label{WFHQV}
\ee
where $\theta_i$ denotes the azimuthal coordinate of the $i$-th particle on the annulus. The integer $\ell_\si$ is a projection of the angular momentum of the $\si$-th component of the wave function of the pair 
on the symmetry axis; the coefficient 1/2 in the exponent reflects the fact that $\ell_\si$ is the momentum of the pair, and, the case $\ell_\uparr = \ell_\dwnarr$ corresponds to a regular full vortex. As can be seen from the above, state $\Psi_{\rm HQV}$  is obtained from the initial state $\Psi_{\rm ESP}$ by a uniform spin-dependent boost. While there might be doubts that the actual HQV is described by such a simple form, it nevertheless should be considered as a good starting point for a variational analysis.

In the $\bm{d}$-vector formalism $\Psi_{\rm HQV}$ as written above produces a $\bm{d}$ vector which lies  in the plane perpendicular to the spin axis i.e.~to the symmetry axis of the annulus. For an annulus made of single crystal $\rm Sr_2RuO_4$ with the $c$-axis along the symmetry axis, this corresponds to $\bm d$ being in the $ab$-plane of the crystal. Although this configuration is not favored by the spin-orbit interaction, there are theoretical indications that even a very small external magnetic field along $c$-axis can stabilize it \cite{Annett:2007p3637}. Similar considerations also apply to $^3$He-A, and in what follows the in-plane position of the $\bm d$ vector will be assumed. It should be emphasized however that our qualitative conclusion about non-zero spin polarization in HQV does not depend on this assumption.

Knowledge of the state (\ref{WFHQV}) allows one to obtain a variational energy of HQV which can then be minimized to yield its detailed structure. For definiteness from now on we will consider only charged systems. In this case the appropriate for minimization thermodynamic potential is Gibbs energy:
 \be
 	G
	=
	\la {\cal H} \ra
	+
	\int \!\!  d^3\vrr \,
	\Big\{
	\frac 1 {8 \pi}
	\vB^2
	-
	\frac 1 {4\pi}
	\vH \cdot \vB
	\Big\},
\label{G}
\ee
where $\la \cal H \ra$ is the expectation value of the Hamiltonian of the system and $\vB$ and $\vH$ are the magnetic field and induction respectively. For an annulus in a shape of an infinite cylinder with the fields along the symmetry axis $\vH$ is the external magnetic field and $\vB$ is the field inside the cylinder.

As it will be seen below the actual form of $\cH$ is crucial for the stability of the HQV. In the simplest case $\cH$ can be taken to contain only the reduced BCS Hamiltonian $\cH_{\rm BCS}$ with the spin triplet pairing term. However such a choice of $\cH$ combined with the ansatz  (\ref{WFHQV})  never makes HQV thermodynamically stable; at best the HQV, in which $\ell_\uparr \neq \ell_\dwnarr$, is degenerate with a full vortex $\ell_\uparr = \ell_\dwnarr$ at the transition point between states with different vorticities. To lower the energy of HQV below that of a full vortex one needs to account for strong interparticle forces. This can be done in the framework of Fermi liquid theory which is also applicable in the  superconducting state \cite{Leggett:1965}.  With that purpose we write the Hamiltonian of the system as
\be
	\cH
	=
	\cH_{\rm BCS}
	+
	{\cal H}_{\rm FL}.
\label{H}
\ee
Here ${\cal H}_{\rm FL}$ describes energy corrections due to Fermi liquid effects and $\cH_{\rm BCS}$ is a reduced BCS Hamiltonian with spin triplet pairing term representing the weak coupling part of the theory. We  will first evaluate the expectation value of the weak coupling Hamiltonian on the state (\ref{WFHQV}). It can be written as a sum of three terms which have different physical origins:
\be
	E_{\rm BCS}
	=
	E_0
	+
	E_{S}
	+
	T.
\label{EBCS}
\ee
The first term in the equation above is the energy contribution coming from the internal degrees of freedom of Cooper pairs. For the radius of the annulus $R$ much larger than the BCS coherence length $\xi_0$ this contribution will depend on neither the center of mass motion of the Cooper pairs, i.e.~on quantum numbers $\ell_\uparr$ and $\ell_\dwnarr$, nor the magnitude of the magnetic field \cite{Vakaryuk:2008}. Assuming that we are dealing with a big enough annulus this term will not be included in the subsequent considerations.

The second term is the spin polarization energy of the system. Let $N_\si$ be the number of particles with spin projection $\si$. Defining $S$ as a projection of the total number spin polarization on  the symmetry axis
\be
	S \equiv
	(N_\uparr - N_\dwnarr)/2,
\ee
 and $g_{\rm S}$ as the gyromagnetic ratio for the particles in question, the spin polarization energy takes the following form
\be
	E_{S}
	=
	 \frac{(g_S \mu_{\rm B} S)^2}{2 \chi_{\rm ESP}}
	-
	g_S \mu_{\rm B} \, \vB \cdot \vS,
\ee
where $\chi_{\rm ESP}$ is the spin susceptibility of the ESP state calculated in the weak-coupling limit. It should be noted that at this point the total spin polarization $S$ is a variational parameter with the actual value of $S$ to be found by the minimization of energy.

The third term on the r.h.s~of eqn.~(\ref{EBCS}) is the kinetic energy of the currents circulating in the system. Let $\Phi$ be the total flux through the annulus and $\Phi_0 \equiv hc/2e$ be the flux quantum. Introducing the notation
\be
	\ell_{\rm s\Phi} 
	\equiv
	\frac{\ell_\uparr + \ell_\dwnarr} 2
	-
	\Phi/\Phi_0,
	\quad
	\ell_{\rm sp}
	\equiv
	\frac{\ell_\uparr - \ell_\dwnarr} 2,
\ee
one obtains for $T$ the following expression:
\be
	T
	=
	\frac{\hbar^2}{8 m^* R^2}
	\Big\{
	(\ell_{\rm s \Phi}^2 + \ell_{\rm sp}^2)N
	+
	4 \ell_{\rm sp} \ell_{\rm s \Phi}S
	\Big\},
\label{T}
\ee
where $m^*$ is the effective mass of the particles due to Fermi liquid corrections
\footnote{It is implicitly assumed that the system under consideration is Galilean invariant, e.g.~can be adequately described by the jellium model. We ignore complications related to the presence of the periodic lattice potential which are discussed in A.~J.~Leggett, Ann.~Phys.~{\bf 46}, 76 (1968).}. 
In the expression above the first term in the brackets is proportional to the total number of particles $N \equiv N_\uparr + N_\dwnarr$  and is thus fixed for given values of $\ell_{\rm s}$, $\ell_{\rm sp}$. The second term is proportional to the spin polarization $S$ and creates an effective Zeeman field in the HQV state due to a mismatch between velocities of the up and down spin components. The value of this field and hence the magnitude of the thermal equilibrium spin polarization should be found by energy minimization. However, as have already been emphasized, minimization of $E_{\rm BCS}$ (or corresponding Gibbs potential $G$ when self inductance is important) does not produce a stable HQV;  at best the HQV is degenerate with a full vortex at the transition point at which the effective Zeeman field vanishes due to vanishing of $\ell_{\rm s\Phi}$.

To make an HQV stable one needs to go beyond the weak coupling Hamiltonian and introduce strong coupling effects.  This can be done in the framework of Fermi liquid theory, in the way indicated by eqn.~(\ref{H}). For that we need to calculate  the change of the  Fermi liquid energy  $E_{\rm FL}$ caused by the presence of spin and momentum currents in the HQV state. These currents are generated by the spin-dependent boost (\ref{WFHQV}) and can be expressed in terms of spin-up and spin-down quasiparticle distributions. Using the standard formalism of Fermi liquid theory one obtains:
\be
\begin{split}
	E_{\rm FL}
	& =
	\frac 1 2 
	\!\!
	\idos
	\!\!\!\!
	Z_0  
	S^2
	+
	\frac
	{N^{-1} \hbar^2}
	{8m^* R^2}
	\,
	\frac 1 3 \,
	\Big[
	( \ell_{\rm s\Phi}^2 F_1 +\ell_{\rm sp}^2 \frac{Z_1} 4)  N^2
	\\
	&+
	4 (\ell_{\rm sp}^2 F_1 + \ell_{\rm s \Phi}^2 \frac{Z_1} 4)  S^2
	+
	4   \ell_{\rm s\Phi} \ell_{\rm sp} (F_1 + \frac{Z_1} 4)  S N
	\Big].
\end{split}
\label{EFL}
\ee
Here $(dn/d\eps)$ is the density of states at the Fermi level and $Z_0$, $Z_1$ and $F_1$ are Landau parameters \footnote{The commonly used set of parameters $F_l^{a}$, $F_l^{s}$ is related to these by $F_l^{s} = F_l$ and $F_l^{a} = Z_l/4$.}. The first term, proportional to $Z_0$, is the energy cost produced by  a spin polarization and the rest describes Fermi liquid corrections due to the presence of the  currents.

It is worth pointing out that expressions (\ref{T}) and (\ref{EFL}), which describe energy transformation under the spin dependent boost (\ref{WFHQV}), can also be written down in terms of momentum and spin currents and are limiting forms of more general transformation rules given in, e.g.~\cite{Cross:1975}.

Now we are in a position to find the equilibrium spin polarization in the HQV state. Minimizing the energy  $E_{\rm BCS} + E_{\rm FL}$ with respect to $S$ we obtain that in equilibrium
\be
	 S 
	=
	(g_S \mu_{\rm B})^{-1}
	\chi
	{\cal B},
\label{S}
\ee
where $\chi$ is the spin susceptibility of the system which, up to terms of order $\epsilon_{\rm F}^{-1} \hbar^2/ 2m^*R^2$, is the spin susceptibility of the ESP state with Fermi liquid corrections; for $^3$He-A at low temperatures the value of $\chi$ is about 0.37 of the normal state susceptibility \cite{Leggett:1965}. The other quantity of interest in eqn.~(\ref{S}), the Zeeman field $\cal B$, has two contributions:
\be
	{\cal B}
	=
	B
	+
	B_{\rm eff},
\ee
which are the external Zeeman field $B$ and the effective Zeeman field $B_{\rm eff}$ caused by  the presence of spin currents:
\begin{gather}
	B_{\rm eff}
	=
	-
	\frac{\hbar^2 (g_S \mu_{\rm B})^{-1}}{2 m^* R^2}
	\,
	 \ell_{\rm sp}
	\ell_{\rm s\Phi}
	\,
	\Big\{ 1 + F_1/3 + Z_1/12 \Big\}.
\label{Beff}
\end{gather}
In thermal equilibrium the effective field is a periodic function of the total flux $\Phi$ with period $\Phi_0$ and changes its sign at flux values  equal to half-integer number of flux quanta. Since at least some of the constants entering eqn.~(\ref{Beff}) are currently not known for $\rm Sr_2 Ru O_4$ it is not possible to give an accurate prediction of the field's magnitude. It is, however,  of order $ \mu_{\rm B}^{-1} \hbar^2/ 2mR^2 = \Phi_0/\pi R^2$; since the first HQV, if stable, exists at about the same value of the external field, this means that the spin polarization produced by the effective field is comparable to that induced by the external field. It is this phenomenon which may provide additional ways for the experimental detection of HQV. Its signature would be a sawtooth contribution given by eqn.~(\ref{Beff}), to the otherwise linear field dependence of the Zeeman spin polarization.

Taking into account both types of spin polarization and omitting the internal energy contribution (cf.~the discussion after eqn.~(\ref{EBCS})),  the energy of the system $E \equiv \la {\cal H} \ra$ can be written as 
\begin{gather}
	E
	=
	- \frac 1 2 
	\chi
	{\cal B}^2
	+
	\frac{\hbar^2 N}{8 m R^2}
	\left\{
	\ell_{\rm s \Phi}^2
	+
	 \ell_{\rm sp}^2
	 \frac{1 + Z_1/12}{1+F_1/3}
	\right\},
\label{E}
\end{gather}
where $m$ is the bare particle mass related to $m^*$ by the usual relation of Fermi liquid theory. For reasonable values of the external field the contribution of the spin polarization energy given by the first term on the r.h.s. of eqn.~(\ref{E}) relative to the total energy $E$ is of order $\hbar^2 \epsilon_{\rm F}^{-1}/2mR^2$ and thus can be safely ignored for the analysis of the stability of HQV.

The region of stability of the HQV  depends on $(1+Z_1/12)/(1+F_1/3)$ which is a zero temperature value for the ratio of spin superfluid and superfluid densities $\rho_{\rm sp}/\rho_{\rm s}$ \cite{Leggett:1975p283}. The stability criterion  found by direct minimization of (\ref{E}) yields the  condition  $\rho_{\rm sp}/\rho_{\rm s} <1$ \footnote{see e.g.~G.~E.~Volovik, JETP Lett.~{\bf 70}, 792 (1999).}, which is usually fulfilled. However, as has been pointed out by Chung \textit{et al.}~\cite{Chung:2007} the self inductance effect, whose treatment  necessitates the use of the Gibbs potential (\ref{G}) constructed out of the energy (\ref{E}), replaces this condition with a much more stringent one. In particular, for a cylindrical annulus the stability of HQV requires that
\be
	\rho_{\rm sp}/\rho_{\rm s}
	<
	\big(
	1
	+
	Rd/2\lambda_{\rm L}^2
	\big)^{-1},
\label{stability}
\ee
with $\lambda_{\rm L}$ denoting the London penetration depth. The value of $\rho_{\rm sp}/\rho_{\rm s} $ in $\rm Sr_2RuO_4$ is currently unknown, however condition (\ref{stability}) makes the existence of HQV in large rings practically impossible. 

The physical interpretation of the stability condition (\ref{stability}) is quite transparent: While the electromagnetic currents which accompany both the full and the half-quantum vortex are well screened in the bulk, the spin currents which are present only in the HQV are not, producing  an additional energy cost over the full vortex. To mitigate such a cost and make HQV stable one needs to reduce the spin current energy by  reducing either spin superfluid density $\rho_{\rm sp}$ or the ``effective volume'' $Rd/2\lambda_{\rm L}^2$ over which the spin currents flow such that the condition (\ref{stability}) is satisfied.

It is further to be remarked that in an annular geometry where one does not have to deal with the vortex core, the stability condition $\rho_{\rm sp}/\rho_{\rm s}< 1$ is still valid for a neutral ESP superfluid such as $^3$He-A. In $^3$He-A the ratio $\rho_{\rm sp}/\rho_{\rm s}$ is known to be well below 1 for all temperatures below critical, hence making the existence of HQV possible in a large part of the phase diagram. By contrast, recent numerical results  \cite{Kawakami:2009} in a solid cylindrical geometry claim that HQV exists only in a high field region and at temperatures sufficiently close to the transition. We believe, however, that the narrowness of the region of HQV stability  obtained in \cite{Kawakami:2009} is due to the omission of the Fermi liquid effects from the consideration.

The real-life question about the thermodynamic stability of HQV is  complicated  and may depend on many factors not included in the preceding discussion. Among these are deviations from the annular geometry, inclusion of spin-orbit interaction and the possibility of the $\bm d$ vector lying in the plane other than $ab$. It should however be emphasized that once the HQV has been stabilized we do not expect our qualitative conclusion about the presence of the spin polarization to be altered by the aforementioned factors since this conclusion originates from one of the defining properties of the HQV, namely velocity mismatch between different spin components. This velocity mismatch shifts  the chemical potentials of up- and down-spin components by an amount of order $\hbar^2/m R^2$ which, in thermal equilibrium, produces an effective spin polarization.

In conclusion, we have shown that the thermal equilibrium state of the half-quantum vortex in the annular geometry should be spin polarized. This effective spin polarization is a periodic function of flux and contributes additively to the spin polarization induced by the external Zeeman coupling. The magnitudes of the two contributions are comparable, thus making the effect potentially observable through, e.g., NMR measurements. This suggests a new way for the experimental detection of half-quantum vortices. 

One of us (V.V.) would like to thank D.~Ferguson for a number of valuable  discussions and C.P.~Slichter for a discussion on the possibility of NMR measurements.  This work was supported by the U.S.~Department of Energy, Division of Materials Sciences under Award No.~DE-FG02-07ER46453 through the Frederick Seitz Material Research Laboratory at the University of Illinois at Urbana-Champaign.


\end{document}